\documentclass[manuscript,screen,review]{acmart}
\usepackage{hyperref}
\usepackage{hyperxmp}
\usepackage{array}
\usepackage{makecell}
\usepackage{xcolor}
\usepackage{ulem}

\AtBeginDocument{%
  \providecommand\BibTeX{{%
    \normalfont B\kern-0.5em{\scshape i\kern-0.25em b}\kern-0.8em\TeX}}}

\setcopyright{acmcopyright}
\copyrightyear{XXX}
\acmYear{XXX}
\acmDOI{XXXXXXX.XXXXXXX}

\acmConference[CSCW'2024]{}{November 9-13,
  2024}{San José, Costa Rica}
\acmPrice{15.00}
\acmISBN{978-1-4503-XXXX-X/18/06}

\usepackage{multirow}

\begin{document}

\title{Application of Prompt Learning Models in Identifying the Collaborative Problem Solving Skills in an Online Task}


\author{Mengxiao Zhu}
\affiliation{%
  \institution{University of Science and Technology of China, Anhui Province Key Laboratory of Science Education and Communication}
  \city{Hefei}
  \country{China}}
\email{mxzhu@ustc.edu.cn}

\author{Xin Wang}
\affiliation{%
  \institution{University of Science and Technology of China}
  \city{Hefei}
  \country{China}}

\author{Xiantao Wang}
\affiliation{%
  \institution{University of Science and Technology of China}
  \city{Hefei}
  \country{China}}

\author{Zihang Chen}
\affiliation{%
  \institution{University of Science and Technology of China}
  \city{Hefei}
  \country{China}}

\author{Wei Huang}
\affiliation{%
  \institution{National Education Examinations Authority}
  \city{Beijing}
  \country{China}}

\renewcommand{\shortauthors}{XXX and XXX, et al.}

\begin{abstract}
Collaborative problem solving (CPS) competence is considered one of the essential 21st-century skills. To facilitate the assessment and learning of CPS competence, researchers have proposed a series of frameworks to conceptualize CPS and explored ways to make sense of the complex processes involved in collaborative problem solving. However, encoding explicit behaviors into subskills within the frameworks of CPS skills is still a challenging task. Traditional studies have relied on manual coding to decipher behavioral data for CPS, but such coding methods can be very time-consuming and cannot support real-time analyses. Scholars have begun to explore approaches for constructing automatic coding models. Nevertheless, the existing models built using machine learning or deep learning techniques depend on a large amount of training data and have relatively low accuracy. To address these problems, this paper proposes a prompt-based learning pre-trained model. The model can achieve high performance even with limited training data. In this study, three experiments were conducted, and the results showed that our model not only produced the highest accuracy, macro F1 score, and kappa values on large training sets, but also performed the best on small training sets of the CPS behavioral data. The application of the proposed prompt-based learning pre-trained model contributes to the CPS skills coding task and can also be used for other CSCW coding tasks to replace manual coding. 
\end{abstract}

\ccsdesc[500]{Human-centered computing~Empirical studies in collaborative and social computing}

\keywords{collaborative problem solving, prompt-based learning, automatic coding, natural language processing}


\maketitle

\section{INTRODUCTION}

Collaborative problem solving (CPS) refers to two or more people working together to solve a problem using their respective skills through information sharing and effective communication \cite{oecd2017pisa}. As one of the most important skills in the 21st century, CPS competence is widely required in many scenarios, including learning environments and the workplace. For example, students are often asked by instructors to work in teams to complete course projects. In the workplace, an increasing number of tasks can no longer be accomplished by individuals but instead require multiple team members to work together. On a global scale, people from different countries collaborate to solve health crises (e.g., COVID-19) and find solutions to other serious global problems (e.g., pollution and global warming) \cite{SayWhat}. In recent years, CPS has attracted significant attention from researchers. Theoretical CPS frameworks have been proposed to conceptualize this abstract construct \cite{oecd2017pisa, ATC21S}, and assessment approaches have also been developed to measure the CPS competence of individuals \cite{oecd2017pisa, DigitalOcean, Psychometrics}. Although CPS competence is important, many students worldwide lack this skill, as reported by the Programme for International Student Assessment (PISA) \cite{oecd2017pisa}. Consequently, it is necessary to deepen the understanding of CPS to improve the CPS abilities of students.

Theoretical frameworks and measurement approaches form the foundation for conducting in-depth CPS analyses. As a composite competence, CPS encompasses various aspects, such as critical thinking, collaboration, communication, and innovation \cite{trilling200921st}. It is also a multimodal, dynamic, and synergistic phenomenon \cite{ouyang2023artificial}, where collaboration and problem solving occur synergistically and influence each other dynamically. Due to the complexity and abstraction of CPS, researchers have proposed theoretical frameworks for conceptualizing this construct and dividing it into multiple subskills (the details are described in Section 2.1) that capture different aspects of CPS competence. 

Two types of assessments are often used for CPS, the traditional multiple-choice methods and simulated-task-based methods. Traditional multiple-choice assessments use text and sometimes images to provide relevant situations to the participants, who then need to choose among the available options to solve problems in imaginary CPS scenarios \cite{fu2023joint}. In this case, the CPS skill levels of individuals can be assessed according to their answers. However, this approach is considered deficient \cite{Ontologies} because it only captures limited information, and more detailed process data on collaboration and problem solving are not available. Hence, researchers have attempted to develop virtual environments to simulate realistic operating spaces \cite{Ontologies} based on the evidence-centered design approach (ECD; \cite{mislevy2003focus}). In a simulated environment, the behavioral trajectories of team members, including their actions (e.g., clicking the mouse and pressing certain keys on the keyboard) and communications (e.g., sending messages to other teammates), can be tracked by a logging system. By analyzing these behavioral data, researchers can further evaluate individuals' mastery level in certain CPS subskills \cite{Ontologies, ThroughComputerGames}, which is beneficial for comprehensively understanding their CPS competence. Such behavioral data can offer insights into how students attempt to achieve goals. However, behavioral data come in very large quantities and are usually unstructured due to the variety of observed behaviors and the inconsistent formats used for data collection. To quantitatively analyze behavioral data, it is necessary to first code the collected behavioral data into specific CPS subskills \cite{CPS-Rater}.  Nevertheless, coding behavioral data is challenging due to the difficulty encountered when attempting to make sense of the large number of operational behaviors of participants. 

Generally, the existing coding approaches can be categorized into two types, manual coding and automatic coding approaches. Manual coding refers to the traditional method of labeling observed data with human coders (e.g., \cite{Ontologies, andrews2020exploring}. The coding process usually involves two or more coders working together on the task based on a mutually agreed-upon coding schema built on a related theoretical framework. After the initial training stage, the coders are expected to reach a high level of agreement regarding the coding results. Because of the rigorous manual coding process, results with high interrater reliability are considered reliable and suitable for further analysis. However, manual coding has limitations. On the one hand, it is a time-consuming and labor-intensive process \cite{SayWhat, ISay}. It is not feasible to rely on manual coding for generating large-scale and real-time codes for CPS behavioral data. On the other hand, if the task scenario or theoretical framework changes, the entire coding process needs to be restarted from scratch. 

To efficiently code CPS behaviors, automatic coding is considered a crucial step toward scaling up the related assessment and learning tasks in the context of CPS \cite{oecd2017pisa, CPS-Rater, PISA2012}. Essentially, the automatic coding task can be regarded as a classification problem, and researchers have begun to explore and develop machine learning and deep neural network models for coding explicit behavioral data \cite{Text-based, GraphAttention, TowardsRobust}. However, traditional classifiers rely on the quantity and quality of the utilized training set to achieve acceptable performance. In real-world applications, a significant amount of high-quality training data may not be readily available. Consequently, the performance of the existing models should be improved, and alternative approaches for building classifiers with limited training data need to be explored.

In this study, we introduce and adopt a prompt learning strategy and develop a prompt-based learning pre-trained model to enable automatic coding without relying on large training sets. Prompt-based learning, as reviewed in \cite{liu2023pre}, is a technique that guides a pre-trained model to generate specific types of outputs by inserting specific prompt texts into the input. By designing effective prompt texts, we can guide the pre-trained model to better understand and handle automatic CPS coding tasks with high accuracy. Furthermore, by leveraging the advantages of prompt-based learning for low-resource data \cite{mahabadi2022perfect, zhou2022prompt, PromptTopic}, we can obtain satisfactory classification results with fewer training data. As a starting point, we explored different combinations of prompt generation methods, including prompt templates, mappings between original labels (i.e., the targeted labels for true CPS subskills), and label words (i.e., the output words downstream of the prompt model), as well as different pre-trained models. The templates and mappings could be generated either manually or automatically. Our results revealed that manually designed templates, along with manually designed mappings, outperformed other prompt generation methods when using the T$5$ \cite{qin2021lfpt5} pre-trained model. We also conducted a comparative analysis with other automatic coding models proposed in previous studies, such as KNN\cite{flor2022towards}, RF \cite{SayWhat}, Linear \cite{BuildingE-rater}, CNN \cite{TextCNN}, LSTM \cite{LSTM}, GRU \cite{GRU}, and pre-trained models based on fine-tuning, such as Finetune-BERT \cite{andrews2022exploring, Bert}, Finetune-RoBERTa \cite{liu2019roberta}, and Finetune-T$5$ \cite{T5}. The results showed that the model developed in our study outperformed the other approaches and achieved the highest accuracy, macro F1 score, and kappa values. Finally, we assessed the performance of our model on small training sets by reducing the amount of input training data and compared the performances of different models with that of our model. The results demonstrated the superior performance of our proposed model on small training sets. In conclusion, this study makes the following contributions.

1)	We introduce a prompt-based learning pre-trained model to address the problem of coding process data in CPS tasks.

2)	The performance of our proposed model is compared with that of other automatic coding models on an empirical dataset derived from a CPS task to demonstrate the superior performance of our model.

3)	By using partial data from the training set, we find that the performance of our model is satisfactory when limited training data are available, and our model outperforms the other models.

\section{RELATED WORK}
In this section, we first review various CPS frameworks proposed in previous research. Next, we provide an overview of the progress made by coding approaches developed for CPS process data, including both manual and automatic coding methods. Finally, we briefly review the existing prompt learning methods.

\subsection{CPS Frameworks}

To comprehend the behaviors and processes of individuals in collaborative problem solving tasks, and to assess their CPS competence, researchers have developed CPS frameworks that can operationalize this complex construct (e.g., \cite{ATC21S, PISA2015}). Most frameworks share a common structure, encompassing both a social aspect related to collaboration and a cognitive aspect related to problem solving \cite{Advancing}. In international CPS assessments, the two most widely used frameworks are the Assessment and Teaching of 21st Century Skills (ATC21s) and the PISA Assessment \cite{Advancing}, and we briefly review these two theoretical frameworks. In addition, we review a third framework \cite{andrews2020exploring} developed for research purposes, which is also adopted in the current study.

As a CPS framework developed for assessment, ATC21s \cite{ATC21S} identifies both social and cognitive aspects. The social aspect covers three components, including \textit{participation}, \textit{perspective taking}, and \textit{social regulation}. \textit{Participation} is a long-term process of becoming a community member, which involves interaction, action, and task completion. \textit{Perspective taking} refers to understanding team members' knowledge, resources, and skills, and responding to others. The final component, \textit{social regulation}, pertains to the strategies and team processes group members employ to facilitate CPS, including negotiating, taking initiative, and assuming responsibility. The cognitive aspect consists of two dimensions, including \textit{task regulation} and \textit{learning and knowledge building}. \textit{Task regulation} is related to problem solving capabilities, such as setting goals, managing resources, exploring problems, and aggregating information. And, \textit{learning and knowledge building} refers to the abilities to plan, execute, reflect, and monitor problem solving.

As a distinct construct, the PISA framework \cite{PISA2015} is composed of four problem solving (also regarded as cognitive) competencies and three collaborative (also regarded as social) competencies. Specifically, the four subdimensions of the problem solving dimension include \textit{exploring and understanding}, \textit{representing and formulating}, \textit{planning and executing}, and \textit{monitoring and reflecting}. And, the three subdimensions of the collaborative dimension comprise \textit{establishing and maintaining shared understanding}, \textit{taking appropriate actions to solve the problem}, and \textit{establishing and maintaining group organization}. These two dimensions interact and cross, forming a matrix with $12$ subskills. 

In addition to the above two frameworks, Andrews and colleagues \cite{andrews2020exploring} also proposed a CPS ontology framework to conceptualize the CPS construct mainly for research purposes. The framework includes nine CPS subskills across two dimensions. The first dimension, the cognitive dimension, involves five subskills, including \textit{exploring and understanding (CEU)}, \textit{representing and formulating (CRF)}, \textit{planning (CP)}, \textit{executing (CE)}, and \textit{monitoring (CM)}. The second dimension, the social dimension, involves four subskills, including \textit{maintaining communication (SMC)}, \textit{sharing information (SSI)}, \textit{establishing shared understanding (SESU)}, and \textit{negotiating (SN)}. In the cognitive dimension, \textit{exploring and understanding} involves actions for exploring problem-related information and building a mutual understanding of the given problem. \textit{Representing and formulating} refers to actions and communication that aim to better visualize problems and form hypotheses. \textit{Planning} concerns communications that are used for the determination of task targets and solutions, as well as subsequent revisions and refinements. \textit{Executing} involves actions and communications during the execution of a task completion plan. Finally, \textit{monitoring} refers to the activities related to determining task completion progress. In the social dimension, \textit{maintaining communication} is about communicating content that is irrelevant to tasks, while \textit{sharing information} is about communicating content that is relevant to tasks. \textit{Establishing shared understanding} refers to group members trying to understand each other’s perspectives. The last subskill, \textit{negotiating}, involves communications used to understand conflicts and propose solutions to reach a consensus. Since the two subskills of \textit{executing} and \textit{monitoring} can occur in either actions or chats, each of them is further split into two components, yielding \textit{executing actions (CE)}, \textit{executing chats (CEC)}, \textit{monitoring actions (CM)}, and \textit{monitoring chats (CMC)}. Thus, the proposed CPS framework includes eleven subskills. The framework provides a theory-driven relationship of the CPS subskills associated with explicit behaviors when participants perform tasks. The empirical analysis conducted in this study used a dataset collected from a three-resistor task \cite{andrews2020exploring}, and this CPS ontology was initially developed for human coding. This CPS framework was thus adopted for building the classifiers. However, our proposed method and model do not depend on any specific CPS framework or any CPS task and can be generalized to other practices and applications beyond CPS.

\subsection{Coding of CPS activities}

To associate conversational and behavioral data from CPS activities with CPS skills, researchers rely on coding methods, which can be roughly divided into manual and automatic coding methods. We first review the manual coding approaches used in the field. Manual coding processes usually depend on coding schemes \cite{flor2022towards}. In general CSCW communities, researchers have developed various manual coding schemes to serve different purposes. For instance, in a study of computer-supported cooperative learning scenarios, \cite{weinberger2006framework} proposed a multidimensional encoding method for dialectical knowledge construction. As another example, \cite{asterhan2009argumentation} developed two complementary encoding schemes with different granularities to annotate dialogs in peer collaboration scenarios. Additionally, \cite{higgins2012multi} coded interactive behaviors such as negotiation and elaboration between different participants. 

With the recent studies on CPS, researchers have also developed coding schemes that fit the CPS simulation environment. For example, \cite{sun2020towards} proposed a hierarchical CPS coding scheme that can effectively capture participants’ behavioral indicators and associate them with CPS skills. \cite{forsyth2020you} proposed an ontology framework for CPS, encoding participants’ chats and actions (e.g., changing their resistance values) into 23 CPS subskills. Moreover, \cite{li2021exploring} proposed a coding scheme combining the 12 CPS subskills classified by the PISA 2015 and students' mastery levels.

In general, the manual coding method is based on a certain theoretical framework that maps a piece of explicit behavior to a specific skill. However, this method has significant limitations. Trained raters need to go through the input data manually, check a large number of corpora, and then map them to specific CPS subskills. Additionally, coders need to ensure rating consistency among themselves, which requires frequent discussions to produce consistent coding results. This process is undoubtedly time-consuming and labor-intensive \cite{SayWhat, ISay}.

With the development of technology, advanced methods have been applied to implement automatic coding, thereby facilitating tasks such as coding text and providing automated feedback \cite{CPS-Rater, zhu2020using}. Recently, researchers in the CPS field have also explored automatic coding approaches. Overall, automatic coding approaches can be divided into two types, i.e., machine learning and deep learning methods. One machine learning method used a linear chain-based conditional random field (CRF) to construct the sequential dependencies of dialog content, and the authors developed an automatic coding system named CPS-rater \cite{CPS-Rater}. This method was proved to be more effective than that of \cite{flor2016automated}, which treated different dialogs independently. In another study \cite{hao2019psychometric}, preselected n-grams and emotions were used to model four aspects of CPS (i.e., sharing ideas, regulating problem-solving activities, negotiating, and maintaining communication). Additionally, the KNN classifier was also used for CPS coding \cite{flor2022towards} and was found to be more satisfactory than naive Bayes classification and comparable to manual encoding.

In addition to the aforementioned automatic encoding methods, \cite{SayWhat,  pugh2022speech} employed a more advanced deep transfer learning approach called bidirectional encoder representations from transformers (BERT) to explore the feasibility of using this model to encode CPS data obtained from simulated indoor environments or real scenes \cite{SayWhat}. They also analyzed the generalizability of several different NLP methods (BERT, n-grams, and word categories) for encoding tasks \cite{pugh2022speech}. As reviewed above, automatic CPS dataset coding is an emerging research direction, and more efficient automatic coding models need to be developed. Additionally, since existing automatic coding models rely on existing manual coding datasets for training, generalization of the existing model requires manually coded datasets. To improve the generalizability of an automatic coding model, we aim to develop a model that depends on a small amount of existing data for automatic coding and can also achieve a relatively high accuracy.

\subsection{Prompt Learning Paradigm}

Before introducing prompt learning, we briefly review the pre-trained language model (PLM) concept, which plays a vital role in facilitating the development of prompt learning methods. When trained on large-scale open corpora \cite{BiasesOfPre-trained}, PLM achieves superior performance in diverse NLP tasks, such as sentiment analysis \cite{pang2002thumbs, socher2013recursive} and machine translation \cite{lopez2008statistical}, due to its ability to embed abundant semantic and syntactic information. Additionally, the model can be adapted to different downstream tasks by learning domain-specific knowledge via fine-tuning \cite{liu2023pre}. Nevertheless, fine-tuning a PLM can be challenging due to the need for large-scale datasets and the involvement of an enormous number of parameters. This challenge is particularly pronounced in low-resource scenarios \cite{PromptTopic}. To address this limitation, a new paradigm called "prompt-based learning", which allows PLMs to process downstream tasks through prompts has emerged \cite{liu2023pre}.

Unlike PLM fine-tuning for a downstream task, the prompt-based method reformulates a downstream task using a textual prompt, effectively turning it into a masked word classification task \cite{BiasesOfPre-trained}. We take text classification as an example. Given the input sentence, "I love this movie", the model is expected to output "positive" or "negative" information about the meaning of the input sentence. However, PLMs designed for text generation cannot directly handle classification tasks. By properly transforming the raw input using the prompt-based method, we can enable text-generated PLMs to perform classification as well. Utilizing the above example, the prompt-based method involves adding a [MASK] token to the input sentence, structuring the input sentence as "I love this movie. It is a [MASK] movie". The model can then generate output words with their associated probabilities, such as "funny", "interesting", or "boring" (referred to as label words). The first two words represent positive emotions, while the third word indicates a negative emotion. The output words can then be mapped to the corresponding emotion words for classification purposes, and this step is known as label-word mapping. Prompt-based models modify the input to adapt a pre-trained model to various downstream tasks, eliminating the need to train a separate model for each task and reducing the requirement for encoding large-scale datasets \cite{liu2023pre, BiasesOfPre-trained}. Therefore, prompt-based learning methods can achieve excellent performance in few- \cite{gu2021ppt, mahabadi2022perfect, cui2022prototypical} and zero-shot \cite{zhong2021adapting, zhou2022prompt, sun2021nsp} tasks.

Due to the advantages of prompt learning, it has garnered increasing attention in recent years. For instance, in the field of mental disease diagnosis, a prompt-based topic-modeling method was developed to detect depression based on question-and-answer data gathered during interviews \cite{PromptTopic}. Researchers utilized the prompt learning paradigm and made topic-wise predictions using the characteristics of the interview data to construct a fusion model for detecting depression. It is worth noting that the sample size of people with mental illness is relatively limited, resulting in even less available data available. Overall, the study demonstrated that the prompt-based model is well suited for addressing the challenge of insufficient training data. The model was also proven to be efficient in personality and interpersonal reactivity prediction tasks. For example, \cite{li2022prompt} employed a prompt-based pre-trained model to participate in a competition involving personality prediction and reactivity prediction, achieving 1st place in both subtasks. The advantage of the designed prompt is that it provides additional personalized information that enhances the performance of the pre-trained model. Furthermore, the prompt-based method has been applied in affective computing. For example, \cite{BiasesOfPre-trained} conducted an empirical study on prompt-based sentiment analysis and emotion detection. They demonstrated the biases of PLMs in prompting by comparing the performances of different prompt templates, label-word forms, and other control variables. This study highlighted the importance of prompt engineering and label-word selections. It is evident from the aforementioned studies that prompt-based models excel in classification tasks and are also effective with small training datasets. In this study, the prompt-based model is applied to automatically code the process data of CPS tasks. Given the pivotal role of the prompt method and pre-trained models in prediction performance, this study aims to determine the appropriate prompt generation method and pre-trained model.

\section{PRELIMINARY ANALYSIS}

In this section, we first introduce the process of collecting and building the utilized dataset, which encompasses participants’ behaviors observed during CPS activities. Next, we present a visualization of the dataset, showcasing the proportions of each subskill (referred to as labels) and the distribution of chat data lengths. Finally, we delve into the data preprocessing steps conducted on the dataset to prepare it for being input into the model.

\subsection{Data Collection}
\paragraph{\textbf{Task.}} The data for this study were collected from an online three-resistor task. The task involves applying relevant physical knowledge of series circuits to adjust resistor values, ensuring that the voltage across the resistance satisfies the requirements of the task. A total of $378$ participants were recruited, and randomly divided into $126$ groups. Each group consisted of $3$ members, with each member responsible for one resistor.

\begin{figure}[h]
  \centering
  \includegraphics[width=10cm]{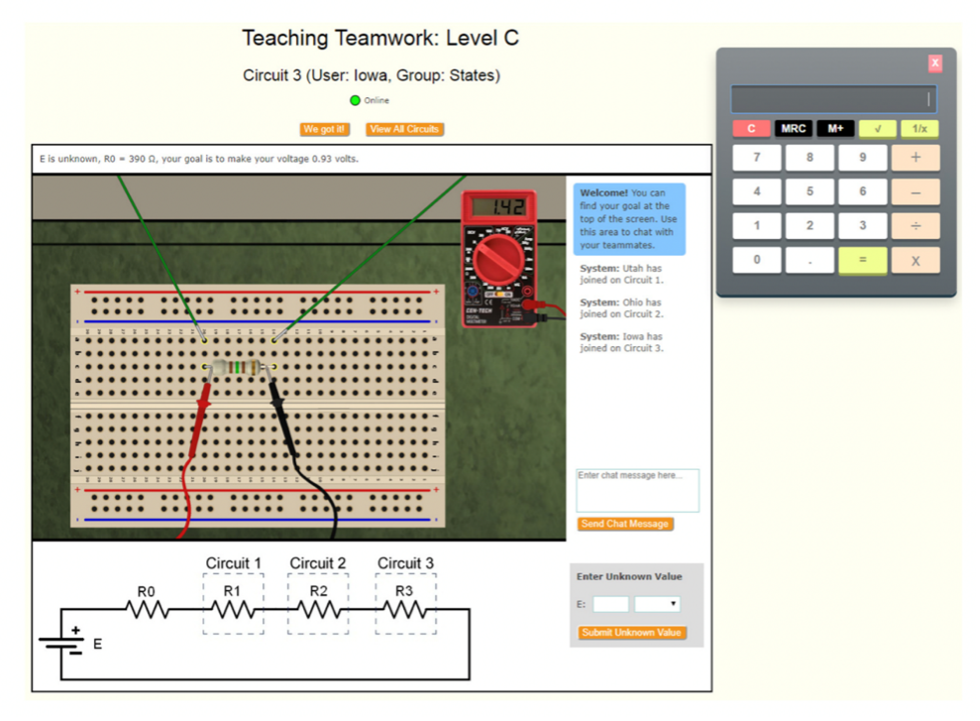}
  \caption{A screenshot of the three-resistor task in the simulated environment.}
  \label{fig:screen}
\end{figure}

The operation interface is represented in Fig.~\ref{fig:screen}. At the top of the screen, the known conditions and targets are displayed. The screen presents a complete circuit structure. The goal of each participant was to adjust the resistor values to reach the target voltage. Because each member received varying information and the resistors in the series circuit influenced each other’s voltage, the group members needed to engage in discussions and collaborate to complete the task. To facilitate communication among the group members, a chat box was provided. Additionally, participants can utilize the calculator in the upper-right corner of the screen for calculations. To accommodate the CPS competence levels of different groups, the task was divided into four levels, primarily differing in the known conditions and goals, as outlined in Table~\ref{tab:condition}.

\begin{table}
  \caption{The condition settings for different task levels.}
  \label{tab:condition}
  \begin{tabular}{cccc}
    \toprule
    
    Task Level&External Voltage (E)&External Resistance (R0)&Goal Voltages\\
    \midrule
    $1$&Known by all teammates&Known by all teammates&Same for all teammates\\
    $2$&Known by all teammates&Known by all teammates&Different for all teammates\\
    $3$&Unknown by teammates&Known by all teammates&Different for all teammates\\
    $4$&Unknown by teammates&Unknown by teammates&Different for all teammates\\
  \bottomrule
\end{tabular}
\end{table}

\paragraph{\textbf{Dataset.}} The data were recorded by a logging system, which included participants' information, such as student IDs and group names, as well as task information and participants' behaviors during the activities. The participants' behavioral data could be classified into two categories. The first category involved manipulating the system, such as changing the resistor or performing calculations, and the second category included chatting with other members in the chat box, such as \textit{"I think it will make it"}, or \textit{"Alright, let's do a big one"}. In total, we collected $50,817$ pieces of data, comprising $15,950$ chat records and $34,867$ manipulation records.

\subsection{Dataset Building}

The collected explicit behavioral data were manually coded by three coders based on the rubric of the CPS ontology framework \cite{andrews2020exploring}. Each record contained information on either an interaction with the simulated task system or a single chat message between team members. For example, the chat message \textit{"we need $6.69V$ across our resistors"} could be classified as planning (CP). The interrater reliability was satisfactory with kappa=$.93$ for the $20\%$ triple-coded samples. Eventually, $50,817$ log entries were classified into $11$ CPS subskills, and the chat data covered $8$ subskills. Table~\ref{tab:example} displays some coding examples. As shown in Table~\ref{tab:example}, the manipulation data could be mapped to a specific subskill since they are generally deterministic. It is more challenging to address chat data due to their diversity and irregularity. To a great extent, chats can be associated with all the subskills, which significantly increases the coding difficulty of the model. Thus, this study focused primarily on automatic chat data coding.

\begin{table}
\centering
\caption{Examples of collected data and their encoding results.}
\label{tab:example}
\begin{tabular}{cccm{4.6cm}c}
\hline
Dimension & Subskill & Label & \multicolumn{1}{c}{Example} & Type \\ 
\hline
\multirow{7}*{Cognitive} & Exploring and Understanding & CEU & Unsystematic/non-strategic use of task components or strategy discovery & Manipulate \\
& Representing and Formulating & CRF & "I feel like it’s the same problem when we had to find an unknown voltage source resistance in circuit analysis" & Chat \\
& Planning & CP & "We need 6.69v across our resistors" & Chat \\
& Executing Actions & CE & Engage in the behaviors consistent with the stated plan for the level (e.g., change resistor to the suggested resistance value) & Manipulate \\
& Executing Chats & CEC & "Adjust yours to 300 ohms" & Chat \\
& Monitoring Actions & CM & Click submit (submit values) & Manipulate \\
& Monitoring Chats & CMC & State where you are or team is about the goal state ("I’m good") & Chat \\
\hline
\multirow{4}*{Social} & Maintaining Communication & SMC & Off-topic conversations not related to the task (e.g., trying to determine the group members’ real name) & Chat \\
& Sharing Information & SSI & "I’m on board 1" & Chat \\
& Establishing Shared Understanding & SESU & Request information: "What is your resistance?" & Chat \\
& Negotiating & SN & Express disagreement: "That’s not right" & Chat \\
\hline
\end{tabular}
\end{table}

\subsection{Data Descriptions}

We conducted a fundamental statistical analysis of the subskill categories in the chat data. Specifically, we calculated the proportion of each type and counted the number of related words that appeared in each chat message to better understand its characteristics.

Table~\ref{tab:freq} shows the frequency and proportion of each classified subskill in the chat data. This reveals that the chat data were unevenly distributed across categories. More than $70\%$ of the chat data pertained to social subskills. \textit{Sharing information (SSI)} appeared most frequently, followed by \textit{establishing shared understanding (SESU)}. Conversely, \textit{representing and formulating (CRF)}, and \textit{planning (CP)} were the least commonly used subskills. In conclusion, social subskills are employed more frequently than cognitive subskills during group communication. The uneven proportions of subskills pose challenges when automatically coding models. The model needed to avoid showing a preference for a specific category during the training process. This ensured that even if the overall prediction accuracy was high, its performance in terms of predicting fewer proportion categories was not extremely poor. Thus, we took this factor into account when evaluating the model performance.

\begin{table}
  \caption{The frequency and proportion of each subskill in all chat data.}
  \label{tab:freq}
  \begin{tabular}{ccccccccc}
    \toprule
    Label&CRF&CP&CEC&CMC&SMC&SSI&SESU&SN\\
    \midrule
    Frequency (n)&$356$&$1066$&$1348$&$1193$&$1292$&$6177$&$3317$&$1149$\\
    Proportion (\%)& $2.24$& $6.71$&	$8.48$&	$7.50$&	$8.13$&	$38.85$&	$20.81$&	$7.23$\\
   \bottomrule
\end{tabular}
\end{table}

\begin{figure}[h]
  \centering
  \includegraphics[width=10cm]{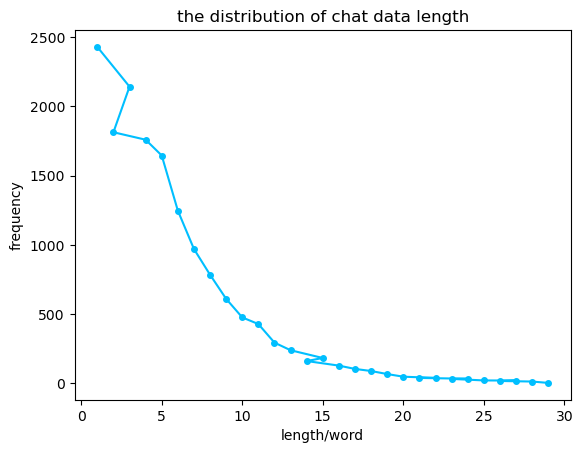}
  \caption{The distribution of chat data length.}
  \label{fig:dis}
\end{figure}

The distribution of the chat data length is depicted in Fig.~\ref{fig:dis}. The distribution exhibited a skewed pattern and was predominantly composed of short sentences. Given that $96.68\%$ of the chat data contained $16$ words or fewer and considering the computational efficiency of the model, we chose a maximum sentence length of $16$ for the subsequent experimental model settings.

\subsection{Data Preprocessing}

To facilitate the subsequent data serialization process, we performed text replacements as outlined in Table~\ref{tab:replacement}. We applied several steps to process the chat data. First, we replaced nouns related to the three-resistor task with special tokens. For example, for the relevant expressions of the four resistors R0-R3 in the circuit, we replaced them with $[R\_zero-R\_three]$ and added them to the pre-trained model. Similarly, we also replaced the expressions of the voltage values, current values, and pure numerical values. In addition, we replaced colloquial abbreviations related to voltage or resistance. Third, expressions referring to team members' nicknames (e.g., tiger, lion) were also substituted with the common names of people to help the language model recognize them as different members of a team.

\begin{table}
  \caption{The replacement rules for some special data during pre-processing.}
  \label{tab:replacement}
  \begin{tabular}{ll}
    \toprule
    Content&	Target\\
    \midrule
    Some numeric values (integer, float)& 	[number]\\
    $R0-R3$, $r0-r3$&	$[R\_zero]$-$[R\_three]$\\
    Some voltage values (integer, float) + V or volts&	[voltage]\\
    Some current values (integer, float) + a or A&	[current]\\
    Volt or voltage&	voltage\\
    R, r, res or resistor&	resistor\\
   \bottomrule
\end{tabular}
\end{table}

\section{METHODOLOGY}

The proposed method consists of a data filtering module and an automatic coding module. The filtering module preprocesses the raw data, which is followed by modeling the input data using two kinds of classification methods based on their categories (chat or manipulation data). This construct is primarily inspired by the design of \cite{andrews2020exploring}. In the automatic coding module, we present a formal formulation and detailed problem descriptions as follows.

\subsection{Problem Formulation} 

In a collaborative problem solving activity, our goal is to predict a CPS subskill Y corresponding to a participant’s explicit behavior X at a certain time.

\subsection{Prompt-Based Coding Method for Chat Data}

\begin{figure}[htbp]  
\centering  
\includegraphics[width=12cm]{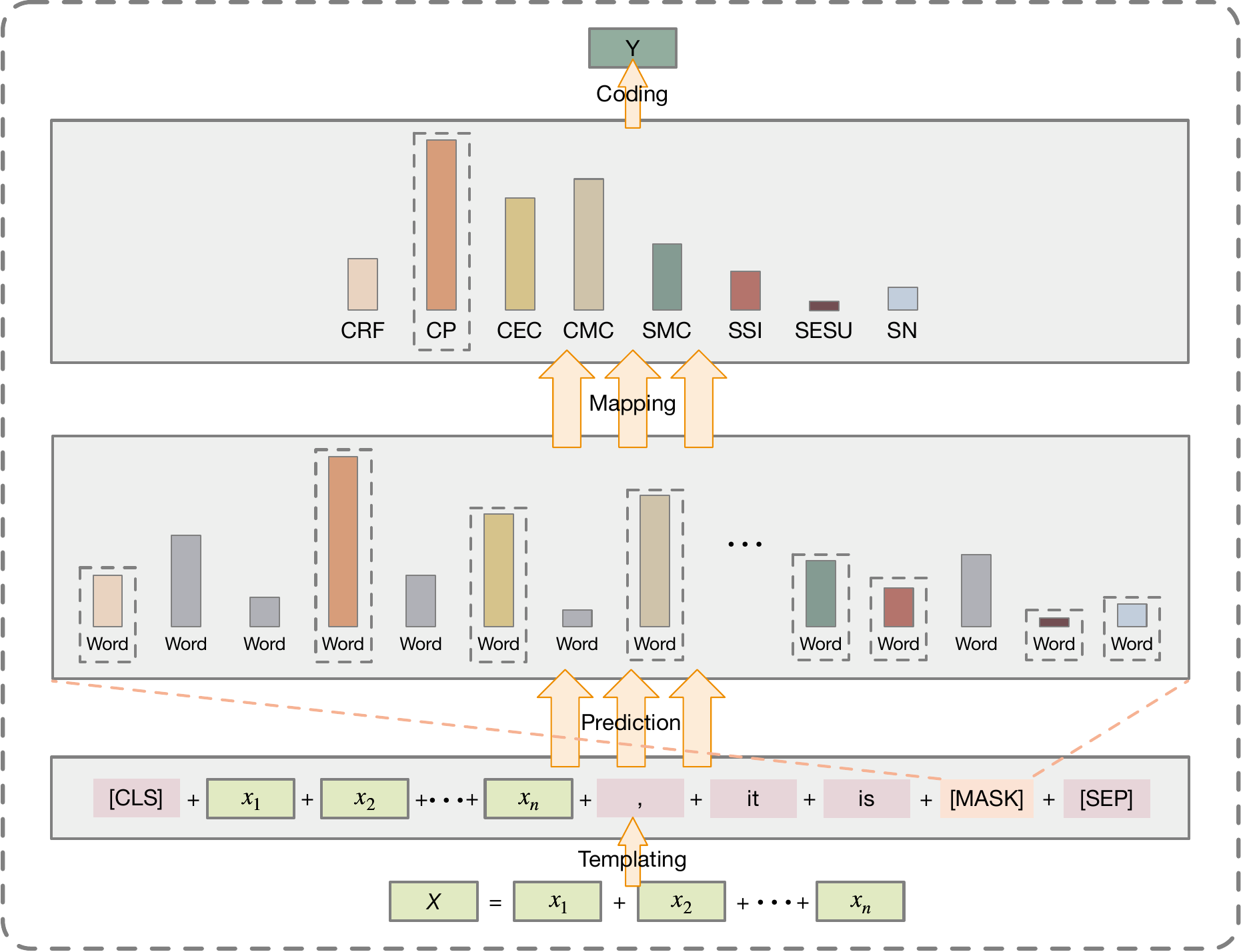}  
\caption{The prompt-based fine-tuning process to code the chat data.}  
\label{prompt}
\end{figure}

We utilized a prompt-based approach to enable the pre-trained language model to automatically code the CPS chat data. This process is illustrated in Fig.~\ref{prompt}. Specifically, for each piece of chat data, we first concatenated it with a manually defined template as follows,

\begin{equation}
  T(X)=[CLS]\,X,\,it\,is\,[MASK]\,[SEP] 
\end{equation}
 
where $T$ represents a modified vector embedding that incorporates the prompt (in this case, the prompt template is “it is [MASK]”); $X$ corresponds to the embedding of the raw chat data; $[CLS]$ and $[SEP]$ denote the beginning and end markers of a sentence in the pre-trained language model, respectively; and $[MASK]$ is the symbol of the position to be predicted by the model. 

After obtaining the templates, we used the pre-trained model to predict the probability of generating each word at the [MASK] position. To elaborate, we constructed a vocabulary $W={w_1,w_2,w_3,\dots,w_n}$, and the probability of generating the word $w_t$ is,

\begin{equation}
    P(w_t) = predict(T(x), PLM, w_t)
\end{equation}

where $PLM$ represents the pre-trained model and $predict(\cdot)$ denotes the use of the $PLM$ to predict the probability of $[MASK]$ belong to $w_t$ in the embedding of $T(x)$. Thus, in this equation, $P(w_t)$ represents the probability of each word $w_t$ (i.e., the word in the vocabulary) being generated in the $[MASK]$ position by the pre-trained model.

After predicting the probability of each word, the model mapped the label words to the original label by calculating the total probabilities of each label word associated with the label. Specifically, if the $s$th label is associated with $k$ words, then the probability of the final automatic coding of the $s$th label is,

\begin{equation}
    P(Y=s) = \sum_i^k{w_i}
\end{equation}

where the prediction probability $P$ is the sum of all probabilities for the associated label words. Ultimately, the result of automatic coding $Y$ corresponds to the label with the highest probability. Thus, the objective function can be defined as follows,

\begin{equation}
    Y = argmax_{s}P(Y=s)
\end{equation}

where $argmax_{s}$ is used to find the argument that maximizes a given function.

\subsection{Rule-Based Coding Method for Manipulation Data}

We use the rule-based model for coding manipulation data. Because the action type was definite, we could code the manipulation data using a one-to-one mapping strategy involving the relevant CPS subskills. For example, actions such as "open Zoom" and "view board in Zoom" were coded as monitoring actions (CM). Another directly corresponding action was "perform calculator with XXX", which could be categorized as executing actions (CE). However, concerning actions that involve changing the value of a resistor from value A to value B, they could be classified as either executing actions (CE) or exploring and understanding (CEU), depending on the group state. If a group already had a plan, the action was labeled as CE. If a group is in an exploring phase, the action is labeled as CEU. Overall, most manipulation data can be directly coded through one-to-one mapping, but some may also require coding techniques based on the specific problem solving stages of the groups during their tasks.

\section{EXPERIMENT AND EVALUATION}

In this section, we present the results of three experiments aimed at demonstrating the advantages of the prompt-based learning pre-trained model in CPS behavioral data classification tasks, especially for cases with small sample sizes. The first experiment focuses on determining the most effective prompt method and pre-trained model combination, in which case the prompt-based learning pre-trained model can achieve superior performance. The second experiment involves a comparative analysis, pitting our model against different classification models proposed in previous automatic CPS coding studies, including both machine learning and deep learning based models. The final experiment aims to verify the superiority of the prompt-based learning pre-trained model in tasks with small sample datasets. We evaluate the performance of the model using accuracy, the macro F1 score, and kappa. The formulas for these metrics are provided in equations $(5)-(7)$ as follows,

\begin{equation}
  Accuracy = \frac{TP+TN }{TP+FP+TN+FN }
\end{equation}

where $TP$ (true positives) denotes the number of correctly classified positive labels; $TN$ (true negatives) denotes the number of correctly classified negative labels; $FP$ (false positives) denotes the number of incorrectly classified positive labels; and $FN$ (false negatives) denotes the number of incorrectly classified negative labels.

\begin{equation}
  Macro\,F1\,score=\frac{F1\,score_{class1}+F1\,score_{class2}+ \ldots+F1\,score_{classN}}{N}
\end{equation}

where $N$ is the number of classes or categories in the classification problem.

\begin{equation}
  Kappa=\frac{P_0-P_e}{1-P_e}
\end{equation}

where $P_0=\frac{TP+TN}{TP+TN+FP+FN}$, and $P_e=\frac{(TP+FP)*(TP+FN)+(TN+FP)*(TN+FN)}{(TP+TN+FP+FN)^2}$.

The former two metrics, accuracy and macro F1 score are commonly used in classification tasks. Given the imbalanced categories of our dataset, we used the macro F1 score to assess the performance of the model. Additionally, we employed kappa to measure the consistency between the results of the model’s coding and manual coding, following the guidelines outlined in \cite{mchugh2012interrater}. A kappa value of $0.60$ indicates acceptable consistency, $0.80$ represents a relatively high level of consistency, and $0.90$ suggests nearly perfect consistency \cite{mchugh2012interrater}.

\subsection{Experiment 1: Comparison Among Different Prompt Methods}

In a prompt method, the selected pre-trained model is crucial to the performance of the resulting model. At the same time, for prompt methods using the same pre-trained model, different training strategies can also lead to significant performance differences. To achieve the best CPS classification performance, we compared four common pre-trained models, namely, BERT\cite{Bert}, RoBERTa\cite{RoBERTa}, GPT-$2$\cite{GPT}, and T$5$\cite{T5}. We designed four different training conditions and described them as follows.

\textit{Manual.} All templates and mappings between the original labels and the label words in the vocabulary are manually defined.

\textit{Trainable Verbalizer (TV).} The mappings between the original labels and label words are determined through training, while the templates are manually defined.

\textit{Trainable Template (TT).} The templates are obtained through training, while the mappings between the original labels and label words are manually defined.

\textit{Trainable Template and Verbalizer (TTV).} Both the templates and mappings between the original labels and label words are obtained through training.

\paragraph{\textbf{Experimental Setup.}} We divide the dataset into a training set, a validation set, and a test set with proportions of $0.70$, $0.15$, and $0.15$, respectively. The number of training epochs is set to $20$, the learning rate is set to $1e$-$5$, and the maximum sentence length is set to $16$. For each model and training environment, we conducted multiple experiments by varying the seeds, which are set to $(0, 1, 2)$. Our model is implemented in PyTorch and trained on an NVIDIA RTX $3090$ GPU device. To effectively evaluate the performance of the model, we use the accuracy, macro F1 score, and kappa values achieved on the test set and calculate the average scores derived from different seeds.

\paragraph{\textbf{Results.}} Table~\ref{tab:result1} summarizes the overall performance of different pre-trained models under various training conditions. From this table, we can observe that under the T$5$-manual condition, the model exhibits the best performance, achieving an accuracy of $0.802$, a macro F$1$ score of $0.725$, and a kappa value of $0.743$, which indicates acceptable consistency with the manual results. Overall, using T$5$ as our pre-trained model with manually defined templates and mappings can yield the best classification results.

\begin{table}
  \caption{Comparison results of different prompt methods.}
  \label{tab:result1}
  \begin{tabular}{ccccc}
    \toprule
    PLM&	Template&	Accuracy&	Macro F1 Score&	Kappa\\
    \midrule

    \multirow{4}*{BERT}& Manual&	0.785&	0.699&	0.719\\
    & TV&	0.791&	0.712&	0.728\\
    & TT&	0.774&	0.702&	0.706\\
    & TTV&	0.784&	0.704&	0.718\\
    \hline
    \multirow{4}*{RoBERTa}& Manual&	0.792&	0.710&	0.733\\
    & TV&	0.800&	0.714&	0.741\\
    & TT&	0.790&	0.706&	0.726\\
    & TTV&	0.790&	\textbf{0.725}&	0.727\\
    \hline
    \multirow{4}*{GPT-2}& Manual&	0.782&	0.695&	0.720\\
    & TV&	0.785&	0.709&	0.722\\
    & TT&	0.771&	0.657&	0.705\\
    & TT&	0.771&	0.658&	0.705\\
    \hline
     \multirow{4}*{T5}& Manual&	\textbf{0.802}&	\textbf{0.725}&	\textbf{0.743}\\
     & TV&	0.792&	0.720&	0.730\\
     & TT&	0.782&	0.714&	0.716\\
     & TTV&	0.780&	0.710&	0.715\\
     
   \bottomrule
\end{tabular}
\end{table}

\subsection{Experiment 2: Comparison with Other Text Classification Models}

Next, we compare the performance of the prompt learning model with that of other text classification models. We select nine baseline text classification models based on previous studies concerning CPS automatic coding, as well as other commonly used text classification models. These baseline models can be classified into three categories, n-gram based methods, deep learning methods, and fine-tuning based methods.

\textit{N-gram based methods.} This class of methods uses an n-gram model to determine the frequencies of word groups and applies TF-IDF for feature engineering to provide input for downstream classification models. For the downstream classification models, we choose linear, K-nearest neighbors (KNN), and random forests (RF) classifiers to perform the final automatic coding task.

\textit{Deep learning methods.} This class of models uses deep learning methods to extract text features and achieves coding via linear neural networks. In the feature extraction stage, we choose the Gated Recurrent Unit (GRU), Long Short-Term Memory (LSTM), and Convolutional Neural Network (CNN) to process the text data.

\textit{Fine-tuning based methods.} Similar to the prompt-based coding method proposed in Section 4.3, this type of method also uses a pre-trained model, with the difference being that these methods directly use a linear neural network to perform automatic coding.

\paragraph{\textbf{Experimental setup.}} The general setup of the experiments remains the same as that described in Section 5.1 but with $85\%$ of the total data as the training set. More setup details regarding the comparison experiments are as follows.

\textit{N-gram based methods.} In the n-gram based methods, we set n to $3$ and the maximum number of features in TF-IDF to $10000$. For the downstream classification models, the setups are as follows. \textbf{1) Linear} uses a two-layer fully connected neural network, and the number of neurons is set to $300$. \textbf{2) KNN} calculates the distances between samples to complete the automatic coding task, and the $K$ value is set to $5$. \textbf{3) RF} uses multiple weak classifiers (decision trees) for automatic encoding, and we set the number of weak classifiers to $10$.

\textit{Deep learning methods.} In the deep learning methods, we set the maximum text length to $20$. The feature extraction methods are set as follows. \textbf{1) GRU} is set to be bidirectional, and the number of layers of hidden layers is set to $2$. Each hidden layer has $256$ neurons. \textbf{2) LSTM} is set in the same way as the GRU, also with $2$ hidden layers consisting of $256$ neurons. \textbf{3) CNN} is set to have $20$ convolutional kernels possessing different sizes, with the sizes of the convolutional kernels ranging from $1$ to $20$.

\textit{Fine-tuning based methods.} We use BERT, RoBERTa, and T$5$ as our pre-trained models and set the maximum text length to $16$.

\begin{table}
\centering
\caption{Comparison results of different classification models.}
\label{tab:result2}
\begin{tabular}{ccccc}
\hline
\multicolumn{2}{c}{Model} & Accuracy & Macro F1 Score & Kappa \\ 
\hline
\multirow{3}{*}{N-gram based methods} & KNN & 0.574 & 0.482 & 0.450 \\
& Linear & 0.621 & 0.487  & 0.498 \\
& RF & 0.714 & 0.615 & 0.619 \\
\hline
\multirow{3}{*}{Deep learning methods} & CNN & 0.659 & 0.524 & 0.543 \\
& GRU & 0.724 & 0.632 & 0.639 \\
& LSTM & 0.608 & 0.453 & 0.469 \\
\hline
\multirow{4}{*}{Finetune} & Finetune-BERT & 0.782 & 0.726 & 0.716 \\
& Finetune-RoBERTa & 0.797 & 0.733 & 0.738 \\
& Finetune-T5 & 0.797 & 0.697 & 0.736 \\
\hline
\multirow{4}{*}{Prompt} & Prompt-BERT & 0.795 & 0.734 & 0.733 \\
& Prompt-RoBERTa & 0.801 & 0.728 & 0.741 \\
&Prompt-T5 & \textbf{0.804} & \textbf{0.743} & \textbf{0.746} \\
\hline
\end{tabular}
\end{table}

\paragraph{\textbf{Results.}} Table~\ref{tab:result2} summarizes the results produced by the prompt-based pre-trained method and the other comparative models, from which it can be seen that our model achieves the best performance concerning all three evaluation criteria. The accuracy is $0.804$, the macro F1 score is $0.743$, and the kappa value is $0.746$. This comparison demonstrates that the proposed classification method based on deep learning is superior to the traditional machine learning methods. Moreover, the methods that use pre-trained models, including prompts and fine-tuning, achieve far better performance than other methods, with our proposed prompt-based model outperforming all other approaches.

\subsection{Experiment 3: Study on Small Training Sets}

In this section, we examine the performance achieved by the prompt model on small samples. We conducted a series of experiments, in which we randomly sampled a portion of the original training set for use as a new training set. Subsequently, we retrained all the models discussed in Section 5.2 using these new training sets and evaluated their performance on the same test set. We employ all three evaluation metrics, namely accuracy, the macro F1 score, and kappa, to comprehensively examine the performance of these models with different training set sizes.

\begin{figure}[htbp]  
\centering  
\includegraphics[width=14cm]{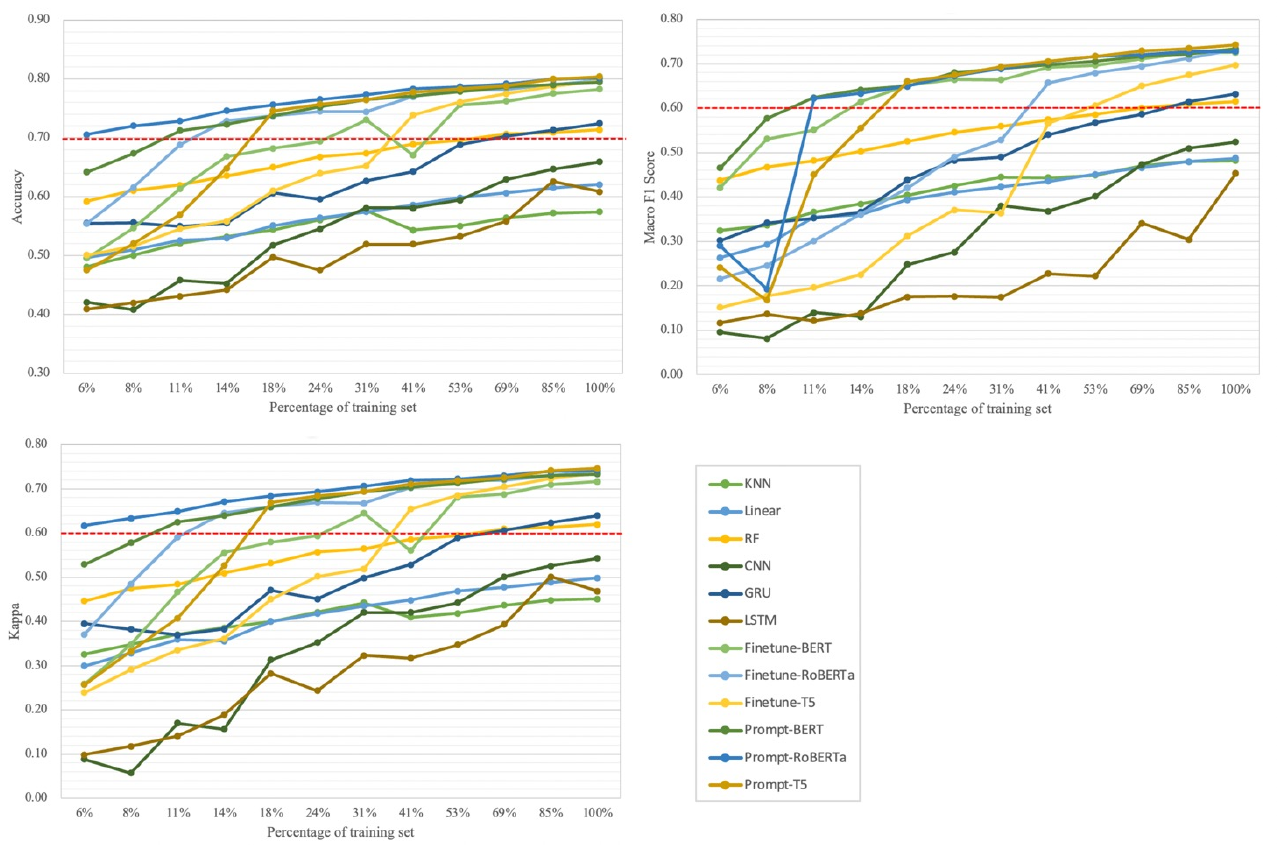}  
\caption{The performance of different models on different scales of training sets, including accuracy, macro F1 score, and Kappa.}  
\label{small_sample}
\end{figure}

\paragraph{\textbf{Experimental setup.}} We randomly sampled various percentages of the original training dataset to create new training sets. Specifically, we used the following percentages to demonstrate the results, $6\%$, $8\%$, $11\%$, $14\%$, $18\%$, $24\%$, $31\%$, $41\%$, $53\%$, $69\%$, and $85\%$. We then retrain all the models discussed in Section 5.2 using these various training set sizes. Subsequently, we tested these retrained models on the original test set and recorded their performance in terms of accuracy, macro F1 score, and kappa. The results of the experiment are shown in Fig.~\ref{small_sample}. Since the accuracy and macro F1 score are the same, we present them in one figure.

\paragraph{\textbf{Results.}} As shown in Fig.~\ref{small_sample}, overall, the prompt-based models (except for Prompt-T5) perform better on the small training samples than do the fine-tuning models. Prompt-RoBERTa and Prompt-Bert are the two best models. To reach the satisfactory predictions indicated by the horizontal red dotted line in the plots, the former model needs only approximately $6\%$ of the original training set (except for the macro F1 score indicator, which needs $11\%$ of the original training set to achieve a satisfactory result), while the latter model needs approximately $11\%$ of the original training set. The Prompt-T5 model needs approximately $16\%$ of the original training set to reach the metric targets. Although this model does not have an obvious advantage on small training sets, it can achieve similar accuracy, macro F1 score, and kappa values to those of the best model, Prompt-RoBERTa, when the training set proportion exceeds $18\%$. For the fine-tuning models, except for Finetune-RoBERTa, which can achieve satisfactory results in terms of the accuracy and kappa indicators with more than $11\%$ of the original training set, Finetune-Bert and Finetune-T5 rely on larger training samples to achieve great results. Additionally, the experimental results demonstrate the significant advantages of pre-trained models. Methods that do not utilize pre-trained models (e.g., GRU or CNN), do not perform as well as the other approaches even when the entire training set is used. In addition, we find that the selected pre-trained model influences the prediction results obtained on small training sets. Specifically, RoBERTa is superior to BERT, and BERT is superior to T5. However, T5 performs better when using the entire training set. 

\section{CODING RESULTS ANALYSIS}
In addition to the accuracy, macro F1 score, and kappa indicators used to evaluate the performance of the automatic coding models, we also performed a confusion matrix analysis and an error analysis to observe the detailed prediction results yielded by the models for every CPS subskill. We employ the two best-performing models in Experiment 3, Prompt-RoBERTa and Prompt-BERT, on the small training set with $11\%$ of the original training set as examples to demonstrate the analysis results.

\subsection{Class Confusion Analysis}
\begin{figure}[htbp]  
\centering  
\includegraphics[width=12cm]{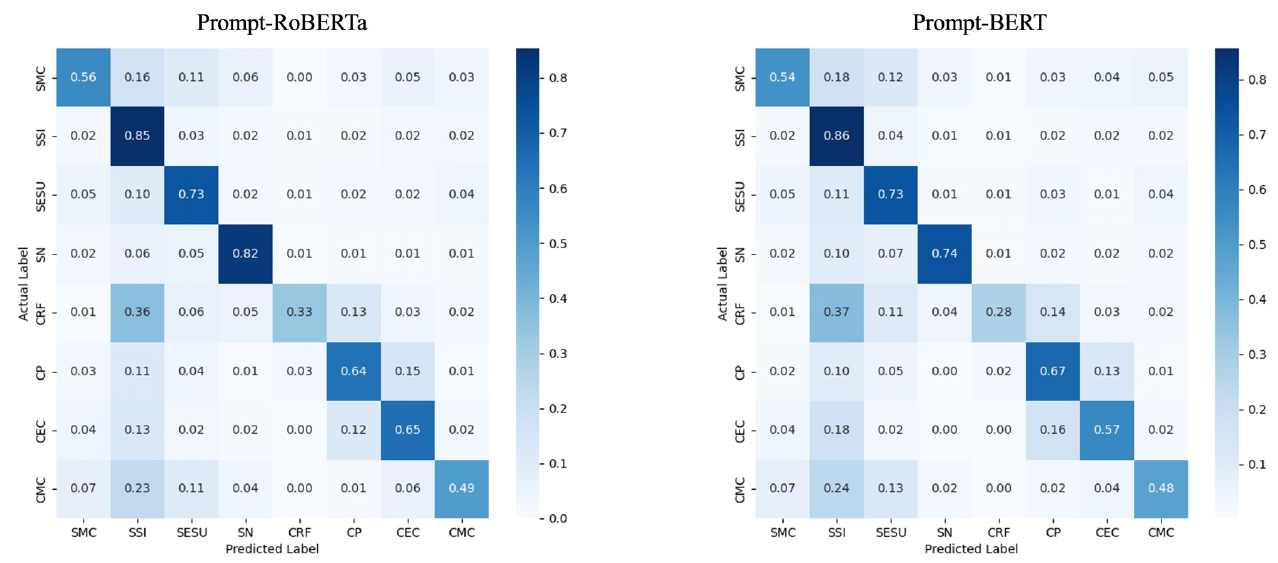}  
\caption{The confusion matrix heat map of the accuracy of Prompt-RoBERTa model and Prompt-BERT model predictions on the $11\%$ of the original training set.}  
\label{heatMap}
\end{figure}

Fig.~\ref{heatMap} represents the confusion matrix heatmaps of the accuracies of the predictions produced by Prompt-RoBERTa (the figure on the left) and Prompt-BERT (the figure on the right) for the eight subskills relative to the actual labels when using $11\%$ of the original training set. The Prompt-RoBERTa model attains the highest prediction accuracy ($0.85$) for the \textit{sharing information (SSI)} subskill, while it has the lowest prediction accuracy ($0.33$) for the \textit{representing and formulating (CRF)} subskill, which is consistent with the subskill frequency results (see Table~\ref{tab:freq}). This model tends to confuse \textit{CRF} with \textit{SSI} ($0.36$) and to confuse \textit{monitoring chats (CMC)} with \textit{SSI} ($0.23$). The reason for this may be that \textit{representing and formulating (CRF)}, and \textit{monitoring (CMC)} chats both involve communication related to tasks, thus requiring the problem or the roles of team members to be understood. Thus, the model may incorrectly regard them as \textit{sharing information (SSI)}. However, \textit{CRF} and \textit{CMC} belong to the cognitive dimension, whereas \textit{SSI} belongs to the social dimension, which shows that the model makes an incorrect prediction in terms of dimensions. Thus, the model may be improved by first considering predicting the high-level dimensions, and then proceeding to more detailed predictions concerning the subskills. The next pair of frequently confused subskills includes \textit{maintaining communication (SMC)} and \textit{SSI} ($0.16$), which may result from the model having trouble determining whether the communication is related to the given task. Overall, the model achieves higher prediction accuracy for the subskills of the social dimension (with all accuracies greater than $0.55$) than for the subskills of the cognitive dimension (with some accuracies lower than $0.50$). The confusion matrix produced by the Prompt-BERT model is similar to that of the Prompt-RoBERTa model. The differences are mainly displayed in the predictions yield for the \textit{negotiating (SN)}, \textit{representing and formulating (CRF)}, and \textit{executing chats (CEC)} subskills. Specifically, the Prompt-BERT model has lower prediction accuracy than the Prompt-RoBERTa model. In conclusion, the prediction results obtained by the two models on the small training set with $11\%$ of the data are both satisfactory. In addition, the results show that the overall prediction accuracy may be improved by improving the prediction accuracies attained for the subskills related to cognition.

\subsection{Error Analysis}
We perform an error analysis to illustrate the errors induced by the Prompt-RoBERTa and Prompt-BERT models. The following lists some cases.

First, we show examples that are incorrectly predicted by the Prompt-RoBERTa model. The chat message of \textit{“click the leads next to the butt plug looking thing”} concerns \textit{planning (CP)}, but the model considers it as \textit{executing chats (CEC)}. Planning generally refers to tasks to be done that have not yet occurred, while executing usually refers to the implementation of a plan. Thus, there is an obvious difference between the timings of these behaviors. However, the model does not detect such a difference, leading to incorrect predictions. Another frequent error occurs between \textit{representing and formulating (CRF)} and \textit{sharing information (SSI)}. Take the chat message \textit{“if mine is [number], the other two r's should sum up to [number]”} as an example. The chat message is labeled as \textit{CRF}, but the model incorrectly classifies it as \textit{SSI}. The word “if” represents a conditional assumption, i.e., an inference made in a certain situation, so this sentence involves representing and formulating, instead of sharing information, which involves sharing something based on existing knowledge. The model may lack the ability to capture keywords for classification purposes.

Second, we show the erroneous cases predicted by the Prompt-BERT model. For instance, the chat message \textit{“I’m on the mark”}, is incorrectly predicted as \textit{sharing information (SSI)} instead of the true label \textit{monitoring chats (CMC)}. On the one hand, this sentence does not include useful information about the task and only aims to inform team members about the progress of their missions. This shows that the model may not accurately capture the information presented in certain cases. On the other hand, the model cannot effectively distinguish between the cognitive and social dimensions. The model regards the chat message of \textit{“lucky guess”}, as \textit{establishing shared understanding (SESU)}, but the true label is \textit{maintaining communication (SMC)}. This message aims to express the joy of making a correct guess and does not include useful information. Thus, the message should be encoded as \textit{SMC}. This example again shows that the model cannot grasp the sophisticated information and message conveyed by the sentence, resulting in an inappropriate prediction.

In conclusion, when the model performs encoding, it cannot take full advantage of the information contained in special words, e.g., conditional words and tense words, to help make more accurate predictions, which may lead to deviations in sentence classification tasks.

\section{DISCUSSION}

Individuals with different skills and knowledge are increasingly required to work on a team to collaboratively solve complex problems \cite{SayWhat}. To better understand and analyze behavioral patterns observed in collaborative activities, recent studies have designed simulated CPS activities to collect process data during tasks. Given the unstructured characteristics of process data, it is necessary to encode and transform them into structured data. However, the existing automatic coding models for CPS process data have relatively low accuracies and significant dependencies on training sets with large numbers of samples. To address this problem, this study proposes an automatic coding model based on prompt learning to label process data, with a primary focus on chat data. In this section, we summarize our main findings from the three conducted experiments and discuss the applications and limitations of this study.

\subsection{Main Findings}

Experiment 1 explores the influences of prompt generation strategies and pre-trained models on the resulting classification performance. We find that manually generating prompts and using T$5$ as the pre-trained model can achieve the best classification results. As noted by \cite{BiasesOfPre-trained}, the classification results may be biased due to the selected prompt and label word forms. Thus, it is suggested that when using prompt-based classification models, the prompt method selection task should be considered. In our study, we compare different prompt methods and find that both the prompt templates and the mappings between the original words and label words display better performance when using the manual design approach. This reflects the uniqueness and complexity of the task of coding CPS process data, making it difficult to obtain appropriate prompt methods only through training. Instead, it is necessary to design suitable prompts according to the activities of the participants in CPS tasks to achieve improved classification performance. Additionally, in our study, we consider different pre-trained models, including BERT and its RoBERTa variant, as well as a transformer-based model (T$5$), and find that T$5$ achieves the best performance. This may be because the T$5$ model has a larger number of model parameters and utilizes a superior relative positional encoding approach instead of the original absolute positional encoding approach. 

Experiment 2 compares our prompt-based learning pre-trained model with nine other widely used classification models. All the models can be divided into four types, namely, n-gram-based methods, deep learning methods, fine-tuning methods, and prompt methods, which share the common feature of considering the task-specific information derived from words in speech \cite{pugh2022speech}. Overall, we find that across the three evaluation indicators, the performance rankings of the different models are as follows, prompt $>$ fine-tuning $>$ deep learning $>$ n-gram. Nevertheless, compared to non-pre-trained model approaches, methods that use pre-trained models perform better because the pretraining task implemented on large-scale data allows them to learn richer and more comprehensive language representations. Furthermore, compared to fine-tuning methods, prompt methods can offer guidance information to the model, helping it better understand the task and domain, ultimately leading to enhanced performance.

Experiment 3 tests the superiority of the prompt-based learning pre-trained model on small training sets. Specifically, based on Experiment 2, we explored the classification results produced by different models with different training set sizes. These results are consistent with our hypothesis, as shown in Fig.~\ref{small_sample}. Prompt-based learning models, such as Prompt-RoBERTa and Prompt-BERT, have the best classification performance. With $85\%$ of the total data as the training set, only $11\%$ of these data ($9.35\%$ of the total data) are needed to achieve satisfactory results. In other words, it is possible to achieve a satisfactory automatic coding effect for a new dataset with only approximately $10\%$ of its data manually coded and used as the training set, which can alleviate the problem of data scarcity \cite{PromptTopic}. Interestingly, we also find that the pre-trained models based on RoBERTa and BERT perform better than T$5$ on small training datasets, and the latter requires approximately $16\%$ of the original training set to reach acceptable performance. However, T$5$ has an advantage when trained on the whole training set.

\subsection{Applications}

The main findings of its work have practical implications. First, performing automatic coding based on a language model can reduce the time and human resources required for manual coding. This strategy relies on only a portion of the manually coded data to learn mapping patterns through training, enabling it to automatically code the remaining data. This method can assist teachers in monitoring group behaviors during CPS activities, enabling them to provide instant support to groups. Additionally, it can help teachers identify students' strengths and weaknesses in cooperative tasks, allowing them to take appropriate actions to improve students' CPS competence levels \cite{pugh2022speech}.

Second, when compared with other models developed in previous research \cite{flor2022towards, andrews2022exploring, pugh2022speech}, our model demonstrates higher classification accuracy and has lower requirements regarding the scale of the input labeled training data. Models based on prompts make pre-trained models directly adaptable to downstream tasks \cite{liu2023pre}. This contextual learning paradigm, allowing a model to learn by providing it with hints, proves to be more effective. This may be the reason why prompt-based learning models outperform the other classification models (e.g., KNN, LSTM, and GRU). \cite{flor2022towards} noted that improving the accuracy of a classifier without increasing the amount of input training data is challenging. Generally, the larger the training dataset is, the better the classification results. However, for CPS tasks, obtaining a large amount of training data is time-consuming and laborious. Our proposed model can achieve satisfactory results with few-shot learning to address this problem.

\subsection{Limitations}
As with all other research, our study has some limitations. The first limitation is that the imposed data preprocessing requirement is relatively high. In different CPS tasks, the task-related words or symbols should be processed differently. Additionally, abbreviations and colloquial words appearing in chat data should also be processed. Due to the diversity of CPS tasks and participants' behaviors, no uniform method is available for preprocessing the task-specific data. Such a method should adapt to the characteristics of the data and model inputs. Another limitation is that our current model considers utterances independently, without accounting for the connections between sentences. Analyzing context can help us more precisely understand the intention of the current utterance, and the mappings between an utterance and the CPS subskills can then be determined more accurately. Finally, the generalizability of the proposed model is not tested in the current study. We only verify the effectiveness of the prompt model on one dataset collected from three-resistor CPS tasks. Therefore, whether the presented findings can be generalized to other CPS tasks and coding scenarios with different CPS frameworks needs to be explored, and this issue will be addressed in our future studies.

\section{CONCLUSION}
This study aims to design an automatic coding model for classifying CPS subskills based on logging data that records participants' explicit behaviors. To achieve this goal, we construct a prompt-based learning pre-trained model and conduct three experiments to verify its superiority. Experiment 1 compares different prompt generation methods and pre-trained models to determine the combination that achieves the best performance. Experiment 2 compares our model with other classification models, while Experiment 3 assesses the performance attained by different models on various small training sets. The results show that our model not only has the highest accuracy, macro F1 score, and kappa values on large training sets but also performs the best on small training sets. Overall, this study demonstrates the effectiveness of the developed prompt-based learning pre-trained model in CPS subskill classification tasks involving low-resource datasets. In the future, we plan to modify our model to implement context-based classification. Additionally, we will test the generalizability of the model to different datasets. We hope that such encoding methods can be extended to more general research fields, such as other human-computer interactions that are commonly studied in the CSCW community, to achieve the text stream, audio stream, and video stream coding.

\bibliographystyle{ACM-Reference-Format}
\normalem
\bibliography{sample-base}

\end{document}